\documentstyle[prd,aps,floats,epsfig]{revtex}  

\begin{document}  
\preprint{astro-ph/0011365} 
\draft 
  
\input epsf 
  
\renewcommand{\topfraction}{0.99} 
\renewcommand{\bottomfraction}{0.99} 
  
\newcommand{\be}{\begin{equation}}
\newcommand{\ee}{\end{equation}}
\newcommand{\ba}{\begin{eqnarray}}
\newcommand{\ea}{\end{eqnarray}}
\newcommand{\rgl}{\rangle}
\newcommand{\lgl}{\langle}
\newcommand{\x}{\mbox{\boldmath $x$}}
\newcommand{\k}{\mbox{\boldmath $k$}}
\renewcommand{\r}{\mbox{\boldmath $r$}}
\newcommand{\bu}{\mbox{\boldmath $u$}}
\newcommand{\nablab}{\mbox{\boldmath $\nabla$}}
\newcommand{\e}{\epsilon}
\newcommand{\de}{\partial}
\newcommand{\nn}{\nonumber \\}
\newcommand{\lm}{{\ell m}}
\newcommand{\lmd}{{\ell' m'}}
\newcommand{\sqtpi}{\sqrt{\frac{2}{\pi}}}
\newcommand{\rhat}{\hat{\r}}
\def \pom {{\hspace{ -0.1em}I\hspace{-0.2em}P}}
\def \GeV {{\rm GeV}}
\def \MeV { {\rm MeV}}
\def \mb {{\rm mb}}
\def \mub {{\rm \mu b}}
\newcommand{\g}{{\tilde g}}
\newcommand{\ag}{\bar{\g}}
\newcommand{\Pl}{{\rm Pl}}


\twocolumn[\hsize\textwidth\columnwidth\hsize\csname 
@twocolumnfalse\endcsname 

\title{Gravitino production in the warm inflationary scenario} 
\author{A.~N.~Taylor$^*$ and Andrew R.~Liddle$^\dag$} 
\address{$^*$Institute for Astronomy, Royal Observatory,
Blackford Hill, Edinburgh, EH9 3HJ, U.~K.\\
$^\dag$Astronomy Centre, University of Sussex, Brighton, BN1 9QJ, U.~K.} 
\date{\today} 
\maketitle 
\begin{abstract}
We estimate the production of gravitinos during and after the end of a
period of warm inflation, a model in which radiation is produced
continuously as the field rolls down the potential producing
dissipation.  We find that gravitino production is efficient for
models in the strong dissipation regime, with the result that standard
nucleosynthesis is disrupted unless the magnitude of the inflaton potential
is very small. Combining this with the constraint from the thermal production
of adiabatic density perturbations we find the dissipation rate must be 
extraordinarily
strong, or that the potential is very flat. 
\end{abstract}

\pacs{PACS number(s): 98.80.Cq, 98.80.-k, 98.80.Es \hfill
astro-ph/0011365}

\vskip2pc] 

\section{Introduction}

The {\em warm inflation} scenario \cite{warm} is an unusual variant on
the inflationary cosmology, in which the inflaton has significant
interactions during the inflationary epoch leading to a continuous
production of radiation. The backreaction of this production on the
inflaton field appears as a viscosity, slowing down the scalar field
evolution and hence aiding slow-roll inflation \cite{YML}. In such a
scenario, inflation can proceed with potentials steeper than those in
standard chaotic scenarios.

Issues concerning the implementation of warm inflation within a
realistic particle physics context have yet to be studied to the same
depth as the standard inflationary scenario \cite{LR,LL}.
Nevertheless, given that the warm inflationary scenario is very
different phenomenologically from the usual picture, it makes good
sense to examine the extent to which its phenomenology is consistent
with observations. The two main purposes of inflation are to provide a
large, nearly homogeneous, patch in the Universe within which
structure formation can take place, and to ensure that unwanted relic
particles do not spoil the successes of the standard hot big bang
cosmology. The first of these has seen a reasonable amount of study
\cite{TB2000}, and so we will consider an example of the latter.

In the context of modern particle physics, the most troublesome relics
are the gravitino and the moduli fields \cite{Sarkar}. We will
consider the gravitino, whose existence arises as the supersymmetric
partner of the graviton, and whose mass is expected to be order of
1~TeV. It is a cosmological threat because if produced in enough
abundance in the early Universe, it is sufficiently long lived to
survive until after nucleosynthesis, at which point its decays spoil
the element abundances \cite{Weinberg}. To avoid this, the ratio of gravitino to
photon number densities must be below about $10^{-12}$. The gravitino
may be produced both by interactions within a thermal bath \cite{kawa} and
by various non-thermal processes \cite{nontherm}. In conventional inflationary
scenarios, the former gives an important upper limit on the reheat
temperature, while the latter may constrain many possible physical
processes.

In this paper we explore the consequences of gravitino production
during and after warm inflation. Warm inflation differs from conventional 
inflation in that radiation is constantly produced during inflation, and the 
radiation density decreases monotonically throughout the evolution, with 
inflation ending when the radiation density overtakes the inflaton energy 
density. There is therefore continuous gravitino production during inflation, 
and also no delay in post-inflationary thermal production due to an intervening 
(p)reheating period. Consequently the gravitino bound is much harder to satisfy. 
We will show that the abundance of gravitinos produced during inflation is 
similar to that produced after inflation, and assess the strength of the 
constraints this imposes on warm inflation model-building.

\section{Evolution of fields during warm inflation}

\subsection{Dissipation effects during inflation}

We review the dynamics of warm inflation closely following Taylor and
Berera \cite{TB2000}, where full details can be found. Warm
inflation is distinguished from ordinary inflation by the presence of
a viscous damping during the inflationary evolution, so that the
inflaton field $\phi$ satisfies the equation
\be
\ddot{\phi} + (3H+\Gamma)\dot{\phi} + V' =0 \,.
\label{evolvphi}
\ee
Here $H\equiv\dot{a}/a$ is the Hubble parameter, $a$ is the
cosmological expansion factor, and $\Gamma$ is the dissipation
coefficient. $V(\phi)$ is the potential of the inflationary field.
For simplicity we assume a spatially-flat universe throughout.  A dot
denotes differentiation with respect to time and a prime with respect
to $\phi$.

The energy density of relativistic species, $\rho_{{\rm rad}}$,
follows from energy conservation as
\be
\dot{\rho}_{{\rm rad}} + 4 H \rho_{{\rm rad}} =  \Gamma \dot{\phi}^2 \,.
\label{evolvgam}
\ee
These equations are completed by the Friedmann equation
\be
H^2 = \frac{8 \pi}{3m^2_{{\rm Pl}}} \left(\rho_\phi + \rho_{{\rm rad}}
	\right) \,,
\label{fried_rad}
\ee
where $m_{{\rm Pl}}$ is 
the Planck mass and the energy density of the inflaton field is 
\be
\rho_\phi = V(\phi) + \frac{1}{2} \, \dot{\phi}^2 \,.
\label{inflden}
\ee
During an inflationary era the potential field dominates both the 
kinetic energy of the inflationary field and the energy density
of the radiation, so the Friedmann equation can be reduced to 
\be
H^2 \simeq \frac{8 \pi}{3m^2_{{\rm Pl}}} V \,.
\label{friedmann}
\ee

Assuming the slow-roll condition, $\ddot{\phi}\ll V'$, the equation of
motion for the inflaton reduces to 
\be
\dot{\phi} \simeq - \frac{V'}{3H(1+r)} \,,
\label{phidot1}
\ee
where 
\be
r \equiv \frac{\Gamma}{3 H} \,,
\ee
is a dimensionless dissipation coefficient, whose value in general is time
dependent. During the inflationary period
the production of radiation will soon settle into a stable state, where
$\dot{\rho}_{{\rm rad}} \ll \Gamma \dot{\phi}^2$, giving 
\be
\rho_{{\rm rad}} \equiv \frac{\pi^2}{30} \, g_* T^4 = \frac{3}{4} \, 
	r \dot{\phi}^2 \,,
\label{radden}
\ee
where $g_*$ is the number of relativistic degrees of freedom. 
In the Standard Model of particle physics
$g_*=106.75$ for $T\ge 300 \, {\rm GeV}$, while in the minimal
supersymmetric model (MSSM) we shall be assuming here this rises to
$g_*=228.75$ once the temperature is above the mass of the supersymmetric 
particles.

Combining Eqs.~(\ref{fried_rad}), (\ref{phidot1}) and (\ref{radden})
we find
\be
\rho_{{\rm rad}} = \frac{1}{2}
\left[\left(1+\frac{2 \epsilon r}{(1+r)^2}\right)^{1/2} -1\right] 
	\rho_\phi \,,
\label{rad2phiden}
\ee
where
\be
\epsilon \equiv \frac{m_{{\rm Pl}}^2}{16 \pi} \left( \frac{V'}{V} \right)^2
\ee
is the usual inflationary slow-roll parameter. 
Eq.~(\ref{rad2phiden}) holds for all values of $r$. In
the limit $r \rightarrow 0$ dissipation is switched off, and
the radiation field vanishes as
\be
	\rho_{{\rm rad}} = \frac{\epsilon r}{2}\,  \rho_\phi \,.
\ee
In this regime, increasing the dissipation factor $r$ increases the
decay of the inflaton field into radiation, while having no effect on
the evolution of the inflaton field.  For fixed $\epsilon$ the fractional 
density of radiation is at a maximum when $r=1$ (equivalently $\Gamma = 3 H$).

\subsection{The strong dissipation regime}
\label{strongdiss}

Our main focus will be the strong dissipation regime, where the
differences from standard inflation are most pronounced.
In the regime of strong dissipation $r\gg 1$ the radiation field is given by
\cite{TB2000}
\be
\label{tempeq}
	\rho_{{\rm rad}} = \frac{\epsilon}{2r}\,  \rho_\phi \,.
\ee
The major effect of increasing the dissipation factor
is to heavily dampen the evolution of the inflaton field, slowing
its evolution down the potential and decreasing the decay into
radiation. 

The conditions for slow-roll and warm inflation to occur are 
\be
	\epsilon < 2 r \,,
\label{const}
\ee
and for an extended period of inflation we need
\be
	|\eta | \ll 3 r^2 \,,
\label{etaeq}
\ee
where 
\be
	\eta \equiv \frac{m_{{\rm Pl}}^2}{8 \pi} \frac{V''}{V} \,.
\ee
Eqs.~(\ref{const}) and (\ref{etaeq}) relax the usual constraints on the
inflationary potential. Supercooled
inflation ends when $\epsilon \approx 1$, while warm inflation takes place 
until
\be
\epsilon \approx 2 r \,,
\ee
when the radiation energy density starts to dominate the energy density
of the inflaton field. At this point the universe makes a smooth
transition from the inflation phase to a radiation-dominated, hot
Friedmann model.

To illustrate the evolution of warm inflation we will consider polynomial 
potentials of the form 
\be
V(\phi) = \lambda m^4 \left( \frac{\phi}{m}\right)^\alpha \,,
\label{pot}
\ee
where we allow $\alpha$ to be a positive real number. It is important to note 
that the only fully complete model of warm inflation, where a period of warm 
inflation comes to a natural end with the field finishing in the minimum of its 
potential, occurs when $\alpha = 2$. For other positive even integers the 
potential has a suitable minimum at the origin. However for $\alpha > 4$ the 
radiation density falls compared to the inflaton energy and so warm inflation 
does not take hold, as can be seen from Eq.~(\ref{tempeq}) since $\epsilon 
\propto \phi^{-2}$ and $r \propto \phi^{-\alpha/2}$ \cite{TB2000}. The 
case $\alpha = 4$ is special in that the 
densities of the two components remain in fixed proportion; inflation proceeds 
forever with the dissipation easing the inflaton asymptotically into the
minimum. Modification to the potential would be required to end inflation. 
We will also consider odd and non-integral values of $\alpha$ 
(taking the field to have positive values). Such potentials have no 
minimum, and indeed may be ill-defined for negative $\phi$; we include them 
merely 
to illustrate the effects of modifications to the slope of the inflaton 
potential. Our range of investigation will cover $1 \le \alpha \le 4$.

For the polynomial potential the slow-roll parameter 
$\epsilon=\alpha^2 m_{{\rm Pl}}^2 /16 \pi \phi^{2}$, and
the number of $e$-folds of expansion to the end of inflation,
$N(\phi)$, is \cite{TB2000}
\be
N(\phi) \equiv \ln \frac{a_{{\rm end}}}{a}
	\approx \frac{\alpha}{4-\alpha} \frac{\rho_\phi}{\rho_{{\rm rad}}} \,,
\label{efold}
\ee
where $\rho_\phi$ and $\rho_{{\rm rad}}$ are the energy densities near
the start of warm inflation once stable radiation production has been
established. If warm inflation starts with an initial stable ratio
$\rho_\phi/\rho_{{\rm rad}}$, it will take $N\sim \rho_\phi/\rho_{{\rm
rad}}$ $e$-folds before the radiation and the vacuum energy are equal.

\begin{figure}
\vspace{-3.cm}
\centerline{\epsfig{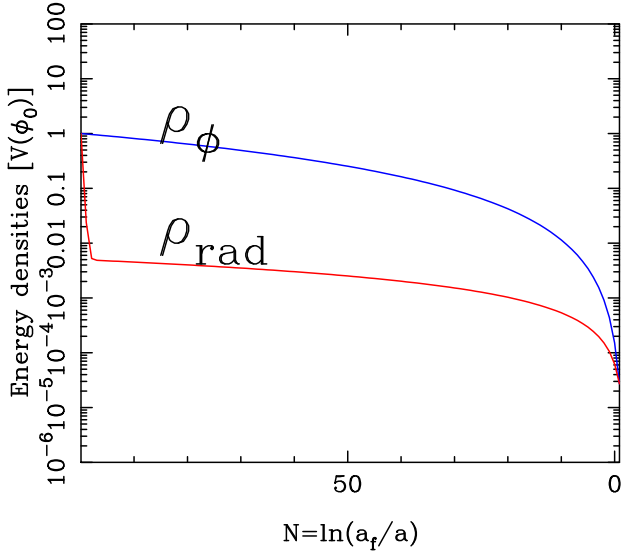}}
\vspace{-3.cm}
\caption{The evolution of energy density of the inflaton and
radiation fields as a function of number of $e$-folds from the end of
warm inflation, for the inflationary potential $V=m^2\phi^2/2$
(see text for details). The energy density of the inflationary
field (upper curves) and the radiation field (lower curves) are taken as 
initially equal, and the vertical scale is arbitrary.}
\end{figure}

Figure 1 shows the evolution of the energy densities of the inflaton
and radiation fields during an inflationary era, when $\alpha=2$. We
numerically solved Eqs.~(\ref{evolvphi}), (\ref{evolvgam}) and
(\ref{fried_rad}) as in Ref.~\cite{TB2000}. The number of $e$-folds is
$N=100$, and we began the model with $\rho_{{\rm rad}} = \rho_\phi$,
although the evolution is insensitive to the initial conditions and
soon settles into its stable configuration. The choice of parameters was made so that the
amplitude of thermally-produced adiabatic perturbations
generated during warm inflation agrees with the amplitude of
temperature fluctuations in the CMB measured by COBE, $\delta_{{\rm H}} = 2
\times 10^{-5}$ (see Section IV.B and Ref.~\cite{TB2000} for details), with
$\Gamma=10^2 m$, $m=10^{-8} m_\Pl$,  and $V^{1/4} \sim 10^{-4} m_\Pl$.

\section{Thermal production of gravitinos}

Gravitinos are too weakly interacting to be able to reach thermal
equilibrium with a radiation bath unless the temperature is around the
Planck temperature. However, although interactions are negligible once
they form, they can be created by two-body processes such as 
\be 
\gamma + \gamma \rightarrow \g + \ag \,.
\ee 
The single-particle decay rate for gravitinos gives a lifetime
of order $m_{3/2}^3/m_{{\rm Pl}}^2$. 

\subsection{Thermal production during warm inflation}

The number density, $n_{3/2}$, of gravitinos produced from the thermal
bath is governed by the equation \cite{kawa}
\be
\dot{n}_{3/2} + 3 H n_{3/2} = \lgl \sigma_{3/2}|v|\rgl 
	n_{{\rm rad}}^2 \,,
\label{evolgrav}
\ee
where 
\be
n_{{\rm rad}} = \frac{\zeta(3)}{\pi^2} \, g_* T^3 
	\approx 0.28 \, g_*^{1/4} \rho_{{\rm rad}}^{3/4} \,,
\label{ngam}
\ee 
is the number density of particles in the thermal bath, and $\zeta(3)
= 1.202$.   Provided 
the typical particle energies well exceed the
gravitino mass, which is always an excellent approximation for us, the
effective total cross-section in a thermal bath for gravitino production, 
including all particle channels, has
been computed as $\sigma_{3/2}^{{\rm tot}} \approx 250/m_{{\rm Pl}}^{2}$
\cite{kawa,fuji}. For convenience we have defined a mean creation rate per 
particle species as $\sigma_{3/2} \approx 250/g_*^2 m_{{\rm Pl}}^{2}$, the 
square arising as the production is a two-body process.\footnote{Our notation 
differs from that of 
Ref.~\cite{kawa}, who define $n_{{\rm rad}}$ to be the number density of a 
single degree of freedom and use the total cross-section.}
We 
will be considering temperatures hot enough that all supersymmetric species 
participate in the thermal bath.
The factor $v\approx 1$ is the velocity of the produced
gravitinos. The gravitino production is  small enough that
backreaction on the radiation density can be neglected, as can
reactions destroying gravitinos by interactions or decays.

After a short period of inflation the gravitino production rate
becomes stable, $\dot{n}_{3/2}\ll \lgl \sigma_{3/2} |v| \rgl
n_{{\rm rad}}^2$, and the gravitino number density is given by
\be 
	n_{3/2} = r_g n_{{\rm rad}} \,,
\label{statgrav}
\ee
where
\be
	r_g = \frac{\lgl \sigma_{3/2} |v|\rgl}{3 H} n_{{\rm rad}} \,,
\label{rg}
\ee
is the dimensionless production rate in units of the Hubble expansion. 
We define the yield of gravitinos as 
\be
Y_{3/2} \equiv \frac{n_{3/2}}{s} \,,
\label{yield}
\ee
where the entropy $s$ is equal to $3.6 n_{\rm rad}$ in the high-energy regime. 
Hence during warm inflation the yield is simply 
\be
Y_{3/2} = \frac{r_g}{3.6} \,.
\label{yield2}
\ee
As long as the stable production hypothesis is valid, the ultimate yield will 
only depend on the situation at the end of warm inflation.

\begin{figure}[t]
\vspace{-3.cm}
\centerline{\epsfig{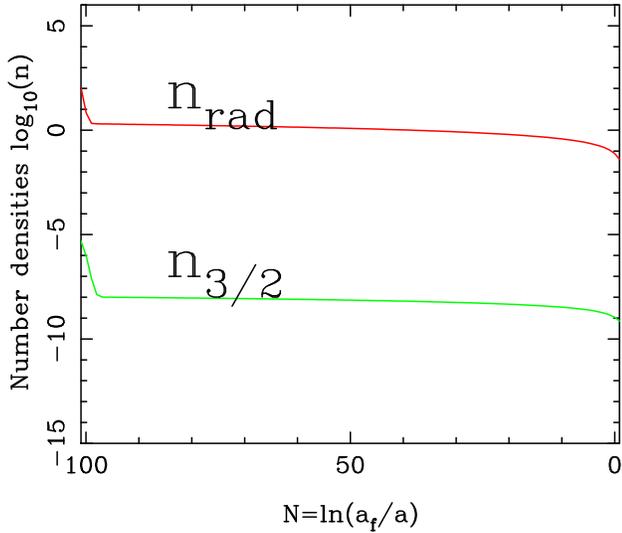}}
\vspace{-3.cm}
\caption{The evolution of number density of the radiation 
and gravitinos as a function of number of $e$-folds from the end of warm 
inflation, for the inflationary potential $V=m^2\phi^2/2$. Number densities are
expressed in units of the radiation number density at the $50^{th}$ $e$-fold.}
\label{fig2}
\end{figure}

Figure~\ref{fig2} shows the evolution of the number densities of the
radiation and gravitino populations during warm inflation. The number densities
are normalised to the number density of radiation at the $50^{th}$ $e$-fold, and
are typically $n_{\rm rad} \sim 10^{-12} m_\Pl^3$, and $n_{3/2} \sim 10^{-20} m_\Pl^3$.
The choice of model parameters is the same as for Figure 1.

The gravitino yield is given by Eqs.~(\ref{ngam}), (\ref{rg}),
(\ref{yield2}), (\ref{tempeq}) and  (\ref{friedmann}) 
\be
Y_{3/2} = 2 \, g_*^{-7/4} \left( \frac{\epsilon}{2 r}\right)^{3/4}
	\, \frac{V^{1/4}}{m_{{\rm Pl}}} \,.
\ee
For a fixed potential, the yield during warm inflation increases with 
increasing potential magnitude and
slope, as there is greater dissipation into radiation, while the yield
decreases with increasing dissipation factor as  strong
dissipation will dampen the decay process.  

For polynomial potentials of the form of Eq.~(\ref{pot})
the yield is given by 
\be
Y_{3/2} = 0.4 \, g_*^{-7/4} \lambda^{5/8} \alpha^{3/2}
	\frac{m}{m_{{\rm Pl}}} \,
	\left(\frac{m_{{\rm Pl}}}{\Gamma}\right)^{3/4} \,
	\left(\frac{\phi}{m}\right)^{(5\alpha-12)/8} .
\label{ypot}
\ee
For this type of potential we see that
$\alpha=12/5$ is a critical slope for gravitino production, 
leading to a constant yield as a function of time. Expressing 
Eq. (\ref{ypot}) in terms of the number of $e$-folds until the
end of warm inflation
\begin{equation}
Y_{3/2} \sim \left[1+\frac{(4-\alpha)}{2\alpha} \,N
	\right]^{(5\alpha-12)/4(4-\alpha)} \,.
\end{equation}
For $\alpha<12/5$ the yield increases as a function of time, or 
$e$-folds, with a maximum at the end of warm inflation.  For
$\alpha>12/5$ the yield is a decreasing function of time.

\begin{figure}[t]
\vspace{-3.cm}
\centerline{\epsfig{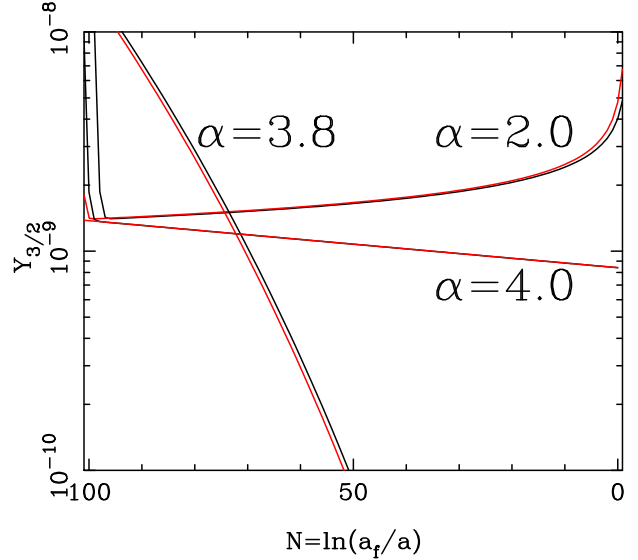}}
\vspace{-3.cm}
\caption{The evolution of gravitino yield for $\alpha=2$, $3.8$ and
$4$. 
The thick lines are calculated numerically, while the lighter
lines are from the analytic expression Eq.~(\ref{ypot}). The warm inflationary 
parameters are chosen to produce the observed amplitude of adiabatic density
perturbations (see text).}
\label{fig3}
\end{figure}

Figure~\ref{fig3} shows the evolution of the yield $Y_{3/2}$ during the warm 
inflation phase, for polynomial potentials
with $\alpha=2$, $3.8$ and $4$.  The solid lines are calculated
numerically, while the lighter lines are from the analytic expression,
Eq.~(\ref{ypot}). Since the gravitino number density quickly settles
into its stable production state, the analytic expression accurately describes 
the evolution.

As the slope is 
increased to $\alpha \rightarrow 4$, the timescale for inflation
is asymptotically stretched out, as discussed in Section~\ref{strongdiss}.
The dependence on $e$-folds asymptotically becomes 
$Y_{3/2} \sim e^{N/4} = (a/a_{\rm end})^{-1/4}$. This weak dependence 
can be seen in Figure 3.

At the end of warm inflation $\e=2r$ and the yield is given by 
\be
Y_{3/2}^{\rm end} = 2 \, g_*^{-7/4} \frac{V^{1/4}}{m_{{\rm Pl}}} \,.
\label{yieldend}
\ee
Hence at the end of warm inflation the yield only depends on 
the magnitude of the potential. The end yield is directly related to the
final temperature, $T_{{\rm end}}$, by 
\be
Y_{3/2}(T_{{\rm end}}) \simeq 1.5 \, g_*^{-3/2} \left( 
	\frac{T_{{\rm end}}}{m_\Pl} \right) \,.
\label{yieldtemp}
\ee

\subsection{Evolution of gravitinos after warm inflation}

Gravitinos have a sufficiently long decay time that they survive
beyond nucleosynthesis. Their decays destroy $^4$He and D nuclei by
photodissociation, and if the gravitino abundance is high enough this
will disrupt the successful predictions of nucleosynthesis. The
important quantity is the ratio of the gravitino to entropy 
densities. 

The entropy is a particularly useful quantity to follow, as the comoving entropy 
is conserved not 
only during normal expansion but also during epochs where species fall out of 
thermal equilibrium and annihilate, changing the number of particle species in 
the thermal bath, so $s \propto 1/a^3$ always.\footnote{An exception to this 
would be if there were particles with late out-of-equilibrium decays.}
The evolution of gravitinos, Eq.~(\ref{evolgrav}), can be rewritten
\be
	\dot{n}_{3/2} + 3 H n_{3/2} = s \dot{Y}_{3/2} = 
	\lgl \sigma_{3/2}|v|\rgl n_{{\rm rad}}^2 \,.
\label{entev}
\ee
The yield produced once radiation domination begins is readily calculated using 
the normal radiation-dominated solution. It is dominated by early time 
production, as is well known, and the total yield from the beginning of 
radiation domination is 
\be
Y_{3/2}-Y_{3/2}^{{\rm end}} \simeq \frac{3 g_*^{-11/6}}{m_\Pl}
	\left[n_{\rm rad}^{1/3}(T_{{\rm end}}) - n_{\rm rad}^{1/3}(T) 
	\right]  \,.
\label{postwi}
\ee
Once $T \ll T_{{\rm end}}$, the right-hand side of Eq.~(\ref{postwi}) equals the 
yield at the end of warm inflation, 
given by 
Eq.~(\ref{yieldend}), since at the end of warm inflation $\rho_\phi \simeq 
\rho_{{\rm rad}}$. We conclude therefore that the production 
of gravitinos during warm inflation can only lead to at most 
a factor of two enhancement 
over the total production at the end of the warm inflation era.

\section{Constraints on warm inflation}

Having calculated the yield of gravitinos during the warm inflationary
era, and shown that the yield does not significantly change afterwards,
we now use this to constrain the warm inflationary parameters. In standard
inflation, there are essentially two free parameters, the amplitude of the
inflaton potential, $V$, and its slope, $\e$. In general these
can be constrained by the gravitino yield and the amplitude of adiabatic
perturbations, the latter being constrained by the observed fluctuations in the 
microwave 
background. In the following section we shall show that the main constraint
from the gravitino production is on the magnitude of the inflaton potential,
while the adiabatic density perturbations constrain the dissipation 
factor $\Gamma$.
We begin with the constraint from the gravitino production.

\subsection{The constraint from nucleosynthesis}

Avoiding overproduction of D + $^3$He constrains the ratio of gravitinos 
to photons at the end of warm
inflation. The details of the constraint depend on the gravitino mass 
\cite{kawa}, 
but for our purposes we can safely adopt a conservative limit
\be
	Y_{3/2} \le 10^{-12} \,.
\ee
Combining this with Eq.~(\ref{yieldend}) for the yield at the 
end of warm inflation, and Eq.~(\ref{postwi}), for the subsequent 
production during radiation domination, we find the following constraint on the
magnitude of the inflation potential 
\be
	V^{1/4} \leq 2 \times 10^{-13} \, g_*^{7/4} m_\Pl.
\ee
This can be expressed in terms of the temperature at the end of 
the warm inflationary phase, via Eq.~(\ref{yieldtemp}), giving 
\be
	T_{{\rm end}} \leq 8 \times 10^{6} g_*^{3/2} {\rm GeV} \,.
\ee
With $g_* = 228.75$ for the MSSM, this gives a constraint in good agreement with 
the standard result \cite{kawa}, though slightly weaker due to our adoption of a 
conservative constraint on $Y_{3/2}$. This is as expected since warm inflation 
has not greatly enhanced the gravitino yield. The main difference in warm 
inflation is how this constraint interacts with other constraints on the 
scenario.

\subsection{The constraint from the amplitude of adiabatic density 
perturbations}

In addition to the yield we can add the independent constraint on 
warm inflation parameters
from the amplitude of perturbations, $\delta^2_{{\rm H}}$, produced from
thermal fluctuations during the warm inflationary era \cite{TB2000},
\be
	\delta_H^2 = 0.57 g_*^{-1/4} \left( \frac{r}{\e} \right)^{3/4}
	\left( \frac{r^2 V}{m_\Pl^4}\right)^{3/4} \,.
\ee
The observational constraint from adiabatic density perturbations does
not actually constrain the amplitude of the potential field alone \cite{TB2000}, 
since
the amplitude is observed at a fixed $e$-fold from the end of warm inflation,
but rather the combination $r^2 V \equiv \Gamma^2 m_\Pl^2/24 \pi$. As we 
usually
assume the amplitude of perturbations is measured at the 50$^{th}$ $e$-fold from 
the
end of warm inflation, some scaling has to be made to match the constraint
from the yield at the end of warm inflation. We may write the number of 
$e$-folds
from the end of warm inflation as
\be	
	N \simeq  A \left(\frac{r}{\e} -1 \right) \,,
\ee
where $A$ depends on the shape of the potential. For polynomial potentials
\be	
	A= \frac{2\alpha}{(4-\alpha)} \,.
\label{Aslope}
\ee
The amplitude of perturbations then constrains the dissipation factor
\be
\Gamma = 10^{-6} 
	\left( \frac{\delta^2_{{\rm H}}}{4 \times 10^{-10}}\right)^{2/3}
	g_*^{1/6}
	\left( 1+ \frac{N}{50 A} \right)^{-1/2} m_\Pl \,.
\label{gammaconst}
\ee

As $r=\Gamma/H$ and $H\sim V^{1/2}$, we can 
combine Eq.~(\ref{gammaconst}) with the constraint on the gravitino yield,
giving a constraint on the dimensionless dissipation rate
\be
r \ge 2 \times 10^{18} 
	\left( \frac{\delta^2_{{\rm H}}}{4 \times 10^{-10}}\right)^{2/3}
	g_*^{-7/3} \left(1+\frac{N}{50A}\right)^{-1/2} .
\label{dissconst}
\ee
This equation is our main result. It shows that if warm inflation is 
to simultaneously produce density perturbations of a satisfactory magnitude and 
avoid overgenerating gravitinos, then the dimensionless dissipation must be 
extremely high. The main reason for this is that high dissipation increases the 
magnitude of density perturbations, allowing the inflationary energy scale to be 
normalized down. With sufficient dissipation, the energy scale becomes low 
enough that the gravitino yield is sufficiently suppressed.

One caveat to Eq.~(\ref{dissconst}) is the factor $A$, relating 
the ratio $r/\e$ to the number of $e$-folds before the end of warm inflation,
when the density perturbations where formed. For polynomial potentials this is 
typically of order unity, but for potentials with very flat slopes, $\alpha 
\rightarrow 0$, the dissipation rate can be 
arbitrarily small, as the radiation production is suppressed.

\section{Conclusions}

A crucial role of inflation is to ensure that unwanted relic particles do not 
survive with abundances capable of spoiling the standard hot big bang model. We 
have studied the production of gravitinos during and after a warm inflationary 
era, and combined the constraint this gives with the requirement that density 
perturbations of the correct magnitude are generated.

We have found that although there is continuous gravitino production during warm 
inflation from interactions in the thermal bath, this does not in itself lead to 
a very significant extra yield of gravitinos over and above that produced at 
the end of warm inflation. Nevertheless, avoiding overproduction of gravitinos 
is much more challenging than in conventional inflationary scenarios, because 
the radiation density is monotonically decreasing throughout the evolution. 
Satisfying the gravitino bound requires that the potential energy at the end of 
warm inflation be very small, $V^{1/4} \lesssim 10^{-9} m_\Pl$. In order for 
density perturbations to have the correct magnitude requires a dissipation 
factor $\Gamma \approx 10^{-6} m_\Pl$, and hence a dimensionless dissipation $r 
\gtrsim 10^{12}$ unless the slope of the potential is extremely flat. Either 
way, it is clear that evading overproduction of gravitinos strongly constrains 
the warm inflation scenario, requiring dimensionless numbers many orders of 
magnitude away from unity.

\section*{Acknowledgments}

A.N.T.~is a PPARC Advanced Fellow, and thanks the University of Sussex Astronomy 
Centre, where this work began, for its hospitality, and Arjun Berera for 
useful discussion.
A.R.L.~acknowledges CITA for hospitality while some of this work was carried
out, and Beatriz de Carlos for discussions on gravitino properties.


\end{document}